\documentclass[sigconf]{acmart}

\usepackage{subfigure}
\usepackage{multirow}
\usepackage{graphicx, caption, subcaption}
\usepackage{enumitem}
\usepackage{array}
\usepackage{booktabs}
\usepackage{bbding}

\renewcommand\footnotetextcopyrightpermission[1]{} 
\acmConference[WWW '25]{Proceedings of the ACM Web Conference 2025}{28 April - 2 May, 2025}{Sydney, Australia}

\newenvironment{shrinkeq}[1] 
{ \bgroup
  \addtolength\abovedisplayshortskip{#1}
  \addtolength\abovedisplayskip{#1}
  \addtolength\belowdisplayshortskip{#1}
  \addtolength\belowdisplayskip{#1}
}
{\egroup\ignorespacesafterend}

\usepackage{titlesec}
\titlespacing*{\subsection}{0pt}{6pt plus 2pt minus 2pt}{2pt plus 2pt minus 2pt}
\titlespacing*{\subsubsection}{0pt}{3pt plus 2pt minus 2pt}{4pt plus 2pt minus 2pt}

\begin{document}

\title{FuXi-$\alpha$: Scaling Recommendation Model with Feature Interaction Enhanced Transformer}

\author{Yufei Ye}
\authornote{Both authors contributed equally to this research.}
\email{aboluo2003@mail.ustc.edu.cn}
\affiliation{%
  \institution{University of Science and Technology
of China}
  \city{Hefei}
  \country{China}
}

\author{Wei Guo}
\authornotemark[1]
\email{guowei67@huawei.com}
\affiliation{%
  \institution{Huawei Noah’s Ark Lab}
  \city{Singapore}
  \country{Singapore}
}

\author{Jin Yao Chin}
\email{chin.jin.yao@huawei.com}
\affiliation{%
  \institution{Huawei Noah’s Ark Lab}
  \city{Singapore}
  \country{Singapore}
}

\author{Hao Wang}
\authornote{Corresponding authors.}
\email{wanghao3@ustc.edu.cn}
\affiliation{%
  \institution{University of Science and Technology
of China}
  \city{Hefei}
  \country{China}
}

\author{Hong Zhu}
\email{zhuhong8@huawei.com}
\affiliation{
  \institution{Consumer Business Group, Huawei}
   \city{Shenzhen}
  \country{China}
}

\author{Xi Lin}
\email{linxi16@huawei.com}
\affiliation{
  \institution{Consumer Business Group, Huawei}
   \city{Shenzhen}
  \country{China}
}

\author{Yuyang Ye}
\email{yeyuyang@mail.ustc.edu.cn}
\affiliation{%
  \institution{University of Science and Technology
of China}
  \city{Hefei}
  \country{China}
}

\author{Yong Liu}
\email{liu.yong6@huawei.com}
\affiliation{%
 \institution{Huawei Noah’s Ark Lab}
 \city{Singapore}
 \country{Singapore}
}

\author{Ruiming Tang}
\email{tangruiming@huawei.com}
\affiliation{%
 \institution{Huawei Noah’s Ark Lab}
 \city{Shenzhen}
 \country{China}
}

\author{Defu Lian}
\authornotemark[2]
\email{liandefu@ustc.edu.cn}
\affiliation{%
  \institution{University of Science and Technology
of China}
  \city{Hefei}
  \country{China}
}

\author{Enhong Chen}
\email{cheneh@ustc.edu.cn}
\affiliation{%
  \institution{University of Science and Technology
of China}
  \city{Hefei}
  \country{China}
}

\renewcommand{\shortauthors}{Ye and Guo, et al.}

\begin{abstract}

Inspired by scaling laws and large language models, research on large-scale recommendation models has gained significant attention.
Recent advancements have shown that expanding sequential recommendation models to large-scale recommendation models can be an effective strategy. 
Current state-of-the-art sequential recommendation models primarily use self-attention mechanisms for explicit feature interactions among items, while implicit interactions are managed through Feed-Forward Networks (FFNs). However, these models often inadequately integrate temporal and positional information, either by adding them to attention weights or by blending them with latent representations, which limits their expressive power. A recent model, HSTU, further reduces the focus on implicit feature interactions, constraining its performance. We propose a new model called \textit{FuXi}-$\alpha$ to address these issues. This model introduces an Adaptive Multi-channel Self-attention mechanism that distinctly models temporal, positional, and semantic features, along with a Multi-stage FFN to enhance implicit feature interactions. 
Our offline experiments demonstrate that our model outperforms existing models, with its performance continuously improving as the model size increases. Additionally, we conducted an online A/B test within the Huawei Music app, which showed a \textbf{4.76\%} increase in the average number of songs played per user and a \textbf{5.10\%} increase in the average listening duration per user.
Our code has been released at \textcolor{blue}{\url{https://github.com/USTC-StarTeam/FuXi-alpha}}.

\end{abstract}


\maketitle

\section{INTRODUCTION}

Recent advancements \cite{gpt3, achiam2023gpt4, kaplan2020scaling, guo2024scaling} in scaling laws have revealed that the performance of Large Language Models (LLMs) systematically improves predictably as the number of model parameters, the volume of training data, and computational resources increase. These findings are crucial as they provide researchers and practitioners with a framework for efficiently allocating limited computational resources to optimize model performance. Building on this foundation, we propose to investigate whether recommendation models also conform to scaling laws. By identifying such models, scaling laws can be utilized to guide the training of larger models, thus enhancing their performance.

Besides the scaling laws found in LLMs such as GPTs \cite{gpt3, achiam2023gpt4}, LLaMAs \cite{touvron2023llama, dubey2024llama}, autoregressive sequential models have been shown to adhere to scaling laws across various domains, including generative image modeling, video modeling, etc \cite{henighan2020scaling}.  
The expansion of Vision Transformers (ViT) has also achieved significant success in the field of computer vision \cite{zhai2022scalingvisiontransformers, dehghani2023scalingvisiontransformers22}.
This revolutionary innovation has also been extended to recommendation models.
Recent studies \cite{zhang2023scaling, shen2024predictive} demonstrate that autoregressive sequence recommendation models also follow these scaling laws. The success of projects like HSTU \cite{hstu, chen2024hllm, wu2024survey} indicates that scaling up sequential recommendation models in accordance with these laws is an effective strategy for developing large-scale recommendation systems.

Sequential recommendation models have been a focal point of research in the field of recommender systems, characterized by a wide array of architectural innovations \cite{xie2024breaking, yin2024dataset, xie2024bridging, shen2024exploring, wangmf, liu2023user,wang2021hypersorec}.
Initially, pooling operations were employed to manage interaction sequences \cite{covington2016deep}. 
With the development of deep learning, more sophisticated models emerged, including CNN-based architectures such as Caser \cite{caser}, GNN-based models like SR-GNN \cite{wu2019session}, RNN-based frameworks like GRU4Rec \cite{hidasi2015session}. Inspired by the huge success of Transformers in NLP, models based on self-attention mechanisms were proposed, leading to notable architectures such as SASRec \cite{kang2018self} and Bert4Rec \cite{bert4rec}.

Besides sequential recommendation models, traditional Deep Learning Recommendation Models (DLRMs), such as DCN \cite{dcn} and xDeepFM \cite{xdeepfm}, also play a crucial role in recommender systems. A fundamental concept in these DLRMs is feature interaction, which is pivotal for enhancing model performance. Feature interactions are categorized into two types: explicit and implicit. Explicit interactions model feature relationships directly through various operators, such as the dot product \cite{fm, dcn}, bilinear functions \cite{xdeepfm}, and attention mechanisms \cite{song2019autoint}. Conversely, implicit interactions are facilitated by applying deep neural networks (DNNs). Although such an approach lacks interpretability, it is extensively used in state-of-the-art DLRMs such as DCN \cite{dcn}, DCNv2 \cite{dcnv2}, DeepFM \cite{deepfm}, and PNN \cite{pnn}. In fact, the integrated DNNs in such models are a key driver of their superior performance. However, previous studies \cite{guo2023embedding, ardalani2022understanding} have indicated that DLRMs do not necessarily exhibit significant performance improvements with increased model size. Nonetheless, the concept of feature interaction can still guide us in designing models.

From the perspective of feature interaction, sequential recommendation models can be conceptualized as exploring the interplay between various features over time. Pooling methods \cite{covington2016deep} have limited expressive capabilities because they overlook the semantic richness of interaction sequences. CNN-based methods \cite{caser} are constrained by a fixed window size, limiting their ability to capture long-range dependencies. RNN-based models \cite{hidasi2015session} interact directly with the previous timestep's hidden state, which can restrict their capacity to model complex interactions. GNN-based approaches \cite{wu2019session} limit feature interactions to directly connected items, thereby narrowing their scope. In contrast, attention-based models, including SASRec \cite{kang2018self}, BERT4Rec \cite{bert4rec}, TiSASRec \cite{tisasrec}, and HSTU \cite{zhai2024actions}, enable comprehensive item interactions. Consequently, these models are more effective at capturing dynamic user interests through interaction sequences. TiSASRec \cite{tisasrec} further improves on SASRec by incorporating time intervals and relative position information, enhancing its performance. HSTU \cite{zhai2024actions} advances this by utilizing positional and temporal information alongside element-wise multiplication to model explicit interactions between items, thereby demonstrating superiority over its predecessors.

\begin{figure}
    \centering
    \setlength{\abovecaptionskip}{0mm}
    \setlength{\belowcaptionskip}{-10px}
    \includegraphics[width=0.8\linewidth]{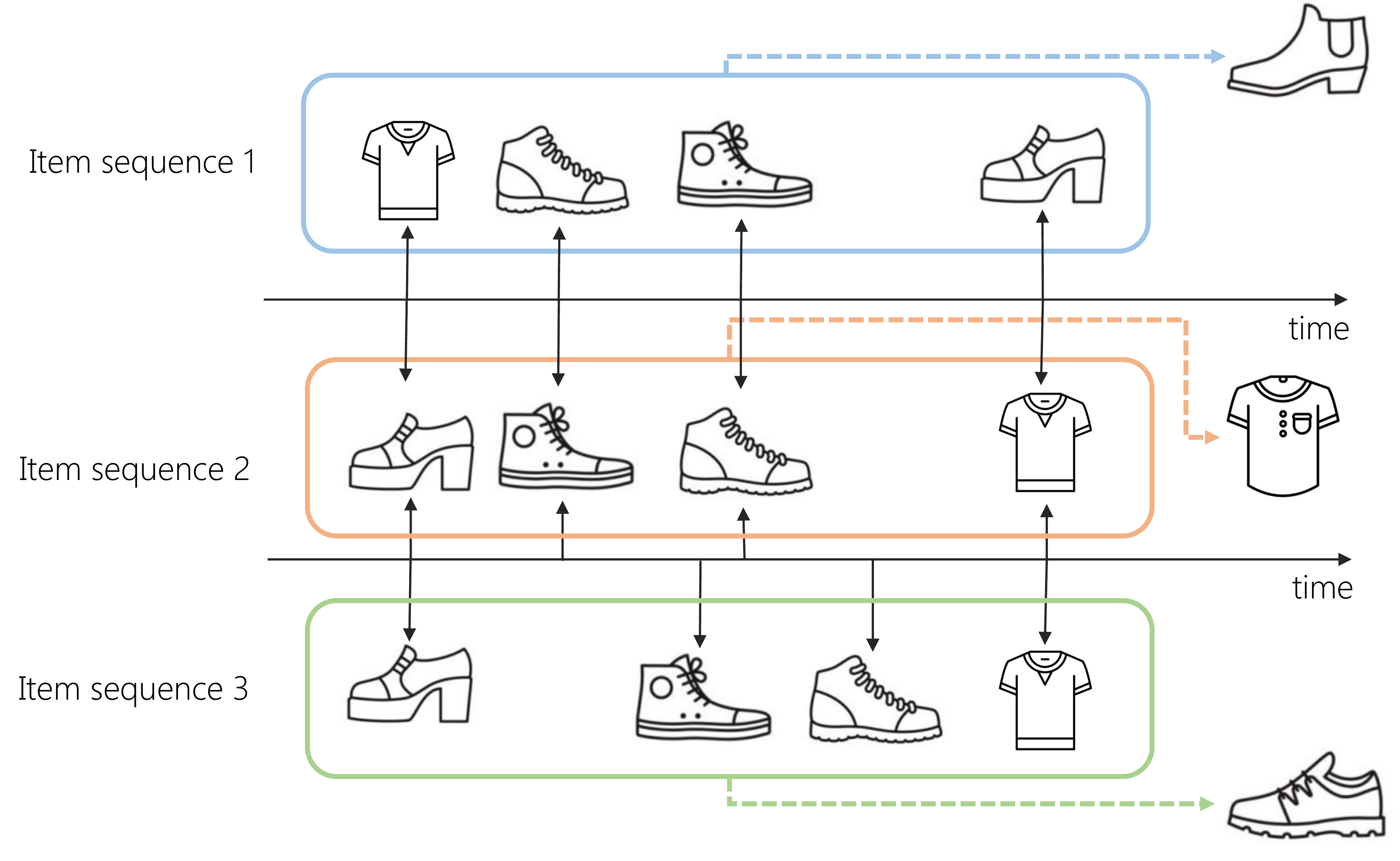}
    \caption{Different temporal intervals or orders between objects may lead to varying subsequent interacted items.}
    \label{fig:intro}
\end{figure}

Despite the significant advancements made in the aforementioned work, there remain several shortcomings that need to be addressed. 
Firstly, previous studies fail to fully leverage temporal and positional information in explicit interactions. They integrate this information by simply adding embeddings to input sequences \cite{kang2018self}, incorporating them into the query and key matrices used in self-attention layers \cite{tisasrec}, or adjusting attention weights \cite{zhai2024actions}. Compared to various methods that facilitate feature interactions, this simple addition lacks expressive capacity. Understanding positional and temporal information is crucial for sequential recommendation because different cues can lead to varying results, as illustrated in Figure \ref{fig:intro}. However, existing models have limited feature interaction with temporal and positional information, hence severely restricting their ability to effectively convey the corresponding temporal and positional cues. Secondly, while HSTU emphasizes explicit interactions, it underemphasizes implicit feature interactions, potentially leading to a loss of nuanced learning processes post-interaction and thus constraining the model's expressiveness.

To address the aforementioned challenges, we propose a novel attention-based model named \textit{FuXi}-$\alpha$. Our approach introduces an Adaptive Multi-channel Self-attention (AMS) layer, which resolves the issue of insufficient feature interactions by modeling the temporal and positional information separately. Furthermore, we integrate a multi-stage feedforward neural network (MFFN) layer to facilitate implicit feature interactions, thereby boosting the model's expressiveness. The proposed method outperforms state-of-the-art sequential recommendation techniques across several benchmark datasets. We also evaluate the model's adherence to scaling laws using a large-scale industrial dataset. The results indicate that performance consistently improves with increased model complexity, highlighting its potential for large-scale recommendation systems. Our contributions are summarized as follows:
\begin{itemize}[leftmargin=*,align=left]
    \item We propose a novel model, \textit{FuXi}-$\alpha$, which adheres to the scaling law by leveraging the perspective of feature interactions.
    \item We design an Adaptive Multi-channel Self-attention (AMS) layer that disentangles the modeling of temporal, positional, and semantic information. We demonstrate that it permits a more expressive representation of temporal and positional information. Additionally, we introduce a Multi-stage Feedforward Network (MFFN) to enhance implicit feature interactions.
    \item We conducted extensive experiments on multiple real-world datasets and online A/B tests on Huawei Music, demonstrating our model's strong performance. Specifically, the online deployment led to an increase of 4.76\% in the average number of song plays per user and a 5.10\% enhancement in the average duration of song playback per user. 
\end{itemize}

\section{RELATED WORK}

\subsection{Scaling Law}

Scaling laws, prevalent in Natural Language Processing (NLP) \cite{gpt3, achiam2023gpt4, kaplan2020scaling, yin2024entropy}, describe the relationship between a model's performance and its size, training data, and computational resources. These laws extend beyond NLP to domains like autoregressive generative models \cite{henighan2020scaling} and visual processing \cite{zhai2022scaling, yang2023lever, yang2024exploring, penglive,  li2024configure}. In the recommendation domain, applying scaling laws is challenging. Studies show that scaling benefits do not always apply to recommendation models \cite{guo2023embedding, ardalani2022understanding}. Issues such as embedding collapse have been reported \cite{guo2023embedding}, and increasing non-embedding parameters in Deep Learning Recommendation Models (DLRMs) offers minimal gains \cite{ardalani2022understanding}.

Despite these challenges, research into scaling laws for recommendation models persists. Studies have explored scaling in user ad activity sequences with generative models \cite{chitlangia2023scaling} and efforts to scale user representation models \cite{shin2023scaling}. A sequential recommendation model with 0.8 billion parameters has been developed, highlighting scaling laws in this domain \cite{zhang2023scaling}. Additionally, it was found that increasing computational resources benefits DLRM less than Generative Recommendations (GR) \cite{zhai2024actions}. This led to the development of HSTU, enhancing the GR paradigm in feature processing, model architecture, and efficiency \cite{zhai2024actions}. 

Our study proposes a model designed to adhere to scaling laws, facilitating its expansion into a large-scale recommendation model for improved performance.

\subsection{Sequential Recommendation}

Sequential recommendation focuses on predicting users' future interests based on past interactions \cite{yin2024learning, han2024efficient, yin2023apgl4sr, han2023guesr}. Early approaches used Markov Chain models \cite{rendle2010factorizing}. With advancements in neural networks, various architectures have enhanced sequential modeling. GRU4Rec \cite{hidasi2015session} uses Gated Recurrent Units to capture sequential data, while Caser \cite{caser} employs CNNs for short-term preference patterns. To model long-term preferences, memory network-based methods \cite{chen2018sequential, huang2018improving, zhang2024learning} were developed. Wu et al. \cite{wu2019session} introduced graph-based interaction modeling. SASRec \cite{kang2018self} and BERT4Rec \cite{bert4rec} leverage self-attention mechanisms for improved recommendations.

In traditional recommendation systems, discriminative-based models typically rank items using a multi-level scoring approach. 
In contrast, generative recommendation models can directly generate the items to be recommended. Following the introduction of HSTU, it has become feasible for autoregressive sequence models that adhere to scaling laws to evolve into generative recommendation models by increasing their model size. HLLM \cite{chen2024hllm} transforms the input IDs into text information encoded by large language models (LLMs), and leverage another LLM for generative sequence recommendation. MBGen \cite{liu2024multi} incorporates behavior tokens into the sequence, thereby improving the model's multi-task capabilities.

In this study, we adopt the autoregressive sequence modeling paradigm to develop a new large-scale recommendation model.

\subsection{Feature Interactions}

Feature interactions play an important role in recommender systems \cite{zhang2022clustering, xu2024multi, zhang2024unified, wang2021hypersorec} and can be divided into explicit and implicit methods.

Explicit interactions are categorized into four types based on their operations: dot product \cite{fm,deepfm,pnn,dcn}, bilinear function \cite{dcnv2,xdeepfm,dcnv3}, convolution \cite{ccpm,liu2019FGCNN,li2019fi-gnn, zhang2019graph, zhang2022hierarchical, zhang2022cglb, zhang2020context, wang2019mcne}, and attention mechanisms \cite{song2019autoint,li2020interhat}. Dot product-based methods like Factorization Machines (FM) and DeepFM extend logistic regression by capturing pairwise interactions \cite{fm, deepfm}. DCN \cite{dcn} models higher-order interactions through product-based cross networks, while DCNv2 \cite{dcnv2} enhances DCN with bilinear functions. DCNv3 \cite{dcnv3} introduces the Exponential Cross Network for more refined modeling. CCPM \cite{ccpm} and FGCNN \cite{liu2019FGCNN} use CNNs for interactions, and Fi-GNN \cite{li2019fi-gnn} applies GNNs. Attention-based methods like AutoInt \cite{song2019autoint} use attention mechanisms, and InterHAt \cite{li2020interhat} employs self-attention for interpretable high-order interactions.

Implicit interactions often use deep neural networks (DNNs) \cite{zhang2016dnn} to simultaneously engage all features. This approach is often combined with explicit interaction structures to enhance overall interaction capabilities. For example, dual-tower architectures like Wide \& Deep and DeepFM integrate low-order explicit interactions with high-order implicit interactions \cite{cheng2016wideanddeep, deepfm}. Models like xDeepFM, DCN, and DCNv2 use DNNs to compensate for certain limitations of explicit feature interactions. \cite{xdeepfm, dcn, dcnv2}. Single-tower structures improve the expressiveness of explicitly crossed features by employing stacked DNNs after explicit interaction structures \cite{he2017nfm, pnn, qu2018pin}.

Inspired by successful feature interaction applications in recommendation models, our work aims to enhance large-scale recommendation models through improved feature interactions.

\section{PROBLEM STATEMENT}

In the domain of sequential recommendation, the primary objective is to predict the next item a user is likely to interact with, based on their historical interaction sequence. Formally, consider a set of users $\mathcal{U} = \{u_1, u_2, \ldots, u_{|\mathcal{U}|}\}$ and a set of items $\mathcal{I} = \{i_1, i_2, \ldots, i_{|\mathcal{I}|}\}$. For each user $u \in \mathcal{U}$, we define an interaction sequence $\mathcal{S}_u = [i_1^{(u)}, i_2^{(u)}, \ldots, i_{n_u}^{(u)}]$, which is a chronologically ordered list of items.

The task of sequential recommendation is to predict the next item $i_{n_u+1}^{(u)}$ that user $u$ will interact with, given the sequence $\mathcal{S}_u$. 
This prediction can be formulated as estimating the probability distribution over the item set $\mathcal{I}$ for the next interaction, conditioned on the historical interactions: $P(i_{n_u+1}^{(u)} = i \mid \mathcal{S}_u)$ for all $i \in \mathcal{I}$.
During training, our objective is to predict the subsequent item $i^{(u)}_{j + 1}$ for every prefix $j$ of the sequence $\mathcal{S}_u$. The desired output sequence is $[i_2^{(u)}, i_3^{(u)}, \ldots, i_{n_u +1}^{(u)}]$ \cite{kang2018self}.

\section{METHODOLOGY}

\begin{figure}
    \centering
    \setlength{\abovecaptionskip}{0mm}
    \setlength{\belowcaptionskip}{-3mm}
    \includegraphics[width=0.4\linewidth]{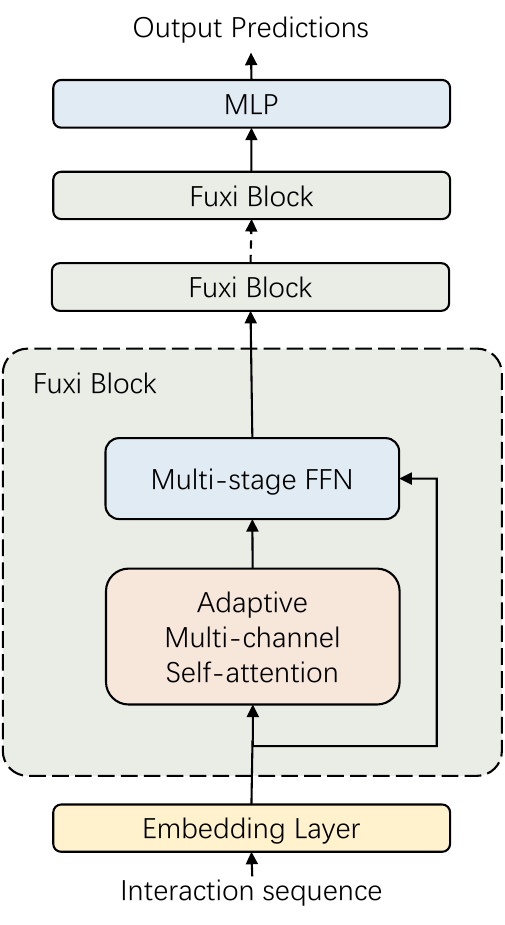}
    \caption{The overall architecture of the proposed \textit{FuXi}-$\alpha$.}
    \label{fig:structure-overview}
\end{figure}

The overview of our model architecture is depicted in Figure \ref{fig:structure-overview}, which is composed of a stack of $b$ \textit{FuXi} Blocks. In the following sections, we will introduce each module individually. Finally, we will discuss the optimization objectives.

\subsection{Embedding Layer}

We convert each user's interaction sequence into a fixed-length sequence of length $n$ through truncation or padding before the embedding layer. Sequences shorter than $n$ are padded with a special "padding item". 
In the embedding layer, each item $i \in \mathcal{I}$ is mapped to a $d$-dimensional vector using a learnable embedding matrix $\mathbf{E} \in \mathbb{R}^{|\mathcal{I}| \times d}$ where $d$ is the latent vector dimensionality. We also employ learnable positional encodings \cite{gehring2017convolutional}, where $\boldsymbol{p}_i$ denotes the positional embedding of the $i$-th position in the sequence. For a user $u$ with a sequence $\mathcal{S}_u = [i_1^{(u)}, \ldots, i_{n_u}^{(u)}]$, the output is $\mathbf{x}^{0} = [\mathbf{e}_1^{(u)} + \boldsymbol{p}_1, \ldots, \mathbf{e}_{n_u}^{(u)} + \boldsymbol{p}_{n_u}, \boldsymbol{0}, \cdots, \boldsymbol{0}]$, where the zero vectors denote the padding items for positions beyond $n_u$ up to $n$.

\subsection{FuXi Block}

The core component of our model is composed of $b$ stacked layers of \textit{FuXi} block which are similar to the transformer decoder \cite{transformer}. Each \textit{FuXi} block consists of an Adaptive Multi-channel Self-attention (AMS) layer and a Multi-stage Feed-Forward Network (MFFN). The adaptive multi-channel self-attention is a variant of the multi-head self-attention \cite{transformer}, while the multi-stage FFN first combines the multi-channel outputs of the AMS layer and then performs implicit feature interactions. 
In this architecture, let $\mathbf{x}^{l-1} \in \mathbb{R}^{n \times d}$ denote the input to the $l$-th layer, and $\mathbf{x}^{l} \in \mathbb{R}^{n \times d}$ denote the output of the $l$-th layer. The initial input for the first layer is given by $\mathbf{x}^{0}$.

\subsubsection{Adaptive Multi-channel Self-attention} 
The AMS layer is designed to effectively capture and utilize the user interest patterns inherent in sequential data. Unlike conventional multi-head self-attention mechanisms, which typically integrate positional encodings directly into the input embeddings, our \textit{FuXi} self-attention separates the processing of hidden states, positional information, and temporal signals into distinct attention heads. This separation allows each head to specialize in capturing different aspects of the sequence data, thereby enhancing the model's capacity to learn complex interest patterns.

\begin{figure}
    \centering
    \setlength{\abovecaptionskip}{5pt}
    \setlength{\belowcaptionskip}{-10pt}
    \includegraphics[width=0.9\linewidth]{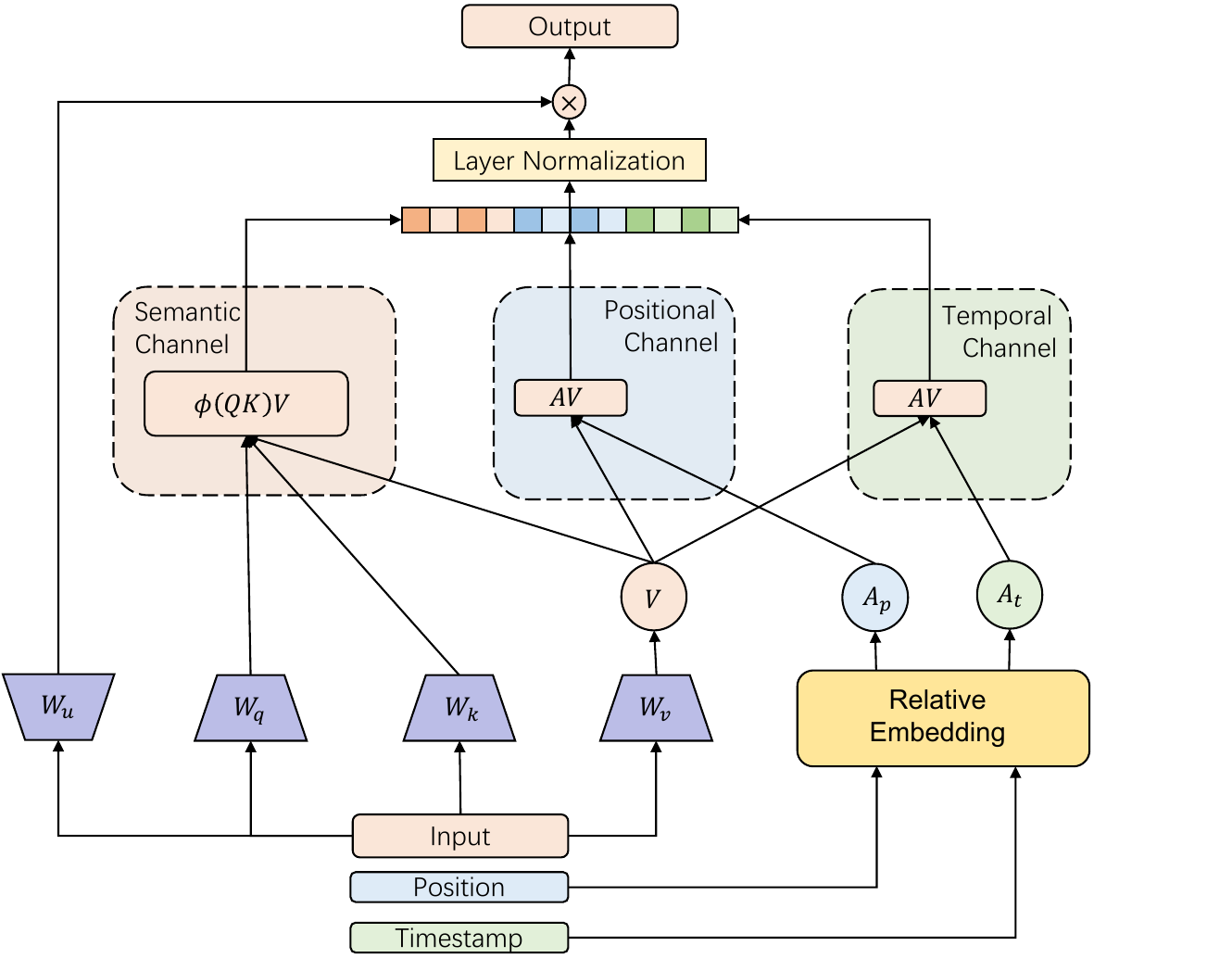}
    \caption{Illustration of Adaptive Multi-channel Self-attention (AMS). In contrast to the conventional multi-head self-attention, AMS decouples the modeling of temporal and positional information from semantics information. }
    \label{fig:structure-fuxi-attention}
\end{figure}

As depicted in Figure \ref{fig:structure-fuxi-attention}, we define three types of channels: semantic, temporal, and positional channels. 
The attention weights in the temporal and positional channels depend only on the difference in relative timestamps and relative positions. 
Additionally, there is no further need to calculate the query and key matrices in these two channels. 
To circumvent the intricacy of the model, we opt not to employ extra value matrices for the temporal and positional heads. Instead, they will share the value matrices with the semantics channel. 
The following approach is used to compute these matrices which is similar to multi-head self-attention:
\begin{align}
    \tilde{\mathbf x}^{l} &= \text{RMSN}(\mathbf x^{l-1}) \\
    \mathbf q^{l} = \phi(\tilde{\mathbf x}^l\mathbf W_{q}^{l}),
    \mathbf k^{l} &= \phi(\tilde{\mathbf x}^l\mathbf W_{k}^{l}), 
    \mathbf v^{l} = \phi(\tilde{\mathbf x}^l\mathbf W_{h}^{l})
\end{align}
where $ \mathbf W_{q}^{l} \in \mathbb R^{d \times d_h}, \mathbf W_{k}^{l} \in \mathbb R^{d\times d_h}, \mathbf W_{v}^{l} \in \mathbb R^{d \times d_h} $ are the learnable parameters. 
RMSN denotes the root mean square (RMS) layer normalization operation \cite{DBLP:conf/nips/ZhangS19a}.
$\phi$ provides nonlinearity which we employ SiLU \cite{elfwing2018silu} here, and $ d_h $ represents the size of each head. The following describes the method for calculating the attention weights for semantic, temporal, and positional channels separately:
\begin{align}
    \mathbf a^{l}_h = \frac{1}{n}\phi(\mathbf q^{l}(\mathbf k^{l})^T), 
    (\mathbf a^{l}_t)_{i,j} = \alpha(t_j - t_i),     (\mathbf a^{l}_p)_{i,j} = \mathbf \beta_{j - i}
\end{align}
where, $\phi$ supplies nonlinearity, and we leverage SiLU once again. Previous studies have demonstrated that the use of SiLU function in self-attention layers outperforms softmax in sequence recommendation tasks \cite{zhai2024actions}. The term $\alpha(t_j - t_i)$ represents the mapping of the difference in timestamps into buckets, where each bucket is associated with a learnable parameter \cite{raffel2020t5}. On the other hand, \(\mathbf{\beta} \in \mathbf{R}^n\) denotes a vector of learnable parameters.

Subsequent to the computation of outputs from the channels, these outputs are concatenated and subjected to RMS layer normalization. 
Following this, the normalized result is element-wise multiplied with the matrix $U$, which is derived from $\tilde{x}^l$. 
The process is encapsulated by the following formula:
\begin{align}
    \mathbf{h}^l &= \text{RMSN}(\text{concat}(\mathbf a^{l}_h \mathbf v^{l}_h, \mathbf a^{l}_p \mathbf v^{l}_p, \mathbf a^{l}_t \mathbf v^{l}_t)) \otimes \phi(\mathbf x^l \mathbf W_u^l)
\end{align}
here $\mathbf W_u^l \in \mathbb R^{d\times 3d_h}$ denotes learnable parameters and $\phi$ denotes SiLU function. We adopted the design of the matrix $U$ in our architecture following HSTU \cite{zhai2024actions} to introduce explicit 2-order interactions. 
For simplicity and clarity, we describe the case with a single head in each channel here. 
However, this approach can be easily extended to multiple heads within each channel, similar to the multi-head self-attention \cite{transformer}.

\subsubsection{Multi-stage Feed-Forward Network} 
\begin{figure}
    \centering
    \setlength{\abovecaptionskip}{0pt}
    \setlength{\belowcaptionskip}{-10pt}
    \includegraphics[width=0.5\linewidth]{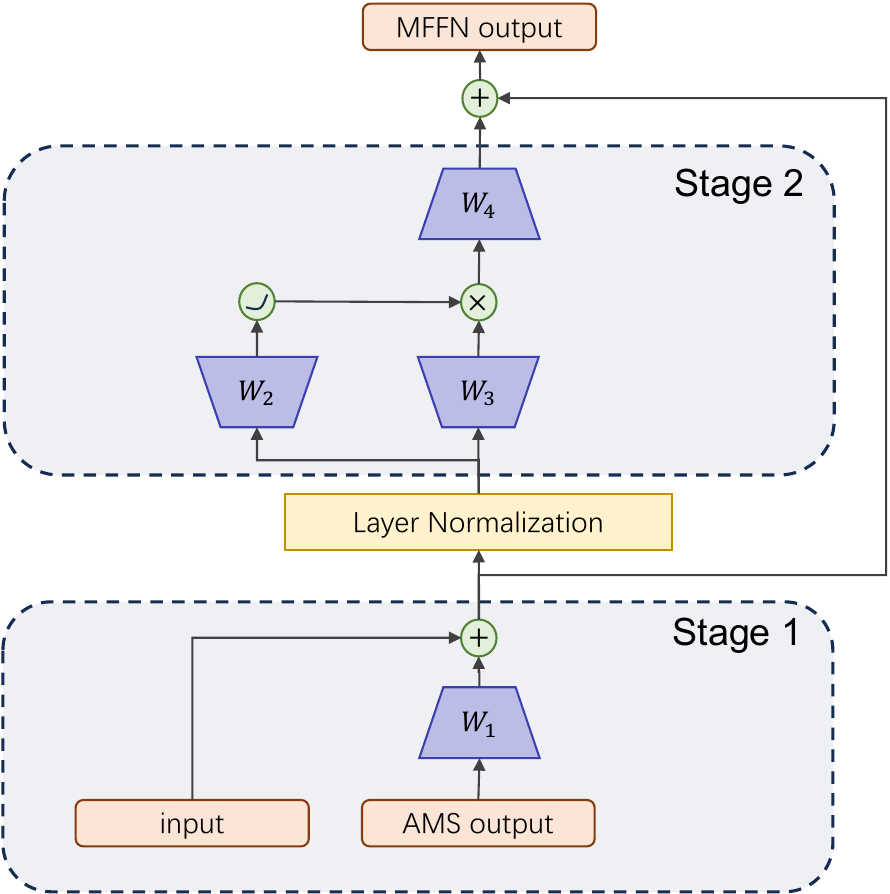}
    \caption{Diagram of MFFN: Stage 1 fuses outputs from different channels; Stage 2 facilitates implicit feature interactions.}
    \label{fig:structure-mffn}
\end{figure}

The MFFN encompasses two distinct stages as depicted in Figure \ref{fig:structure-mffn}. 
In the first stage, the outputs from different channels are fused with the original input of the current layer. Subsequently, in the second stage, implicit feature interactions are conducted.·

In the first stage, MFFN receives the outputs across different channels from the AMS layer and applies a projection transformation characterized by learnable parameters $W_o \in \mathbb R^{3d_h \times d}$. The output of this stage is obtained by combining the projected output with the input of current layer $\mathbf x^l$.
\begin{align}
    \mathbf o^{l} &= \mathbf{h}^l\textbf W_o^{l} + \mathbf x^{l-1}
\end{align}

In the second stage, the primary objective of MFFN is to conduct  implicit interactions. Following LLaMa \cite{touvron2023llama}, we apply RMS layer normalization to the output of the previous stage and followed by a SwiGLU activation \cite{shazeer2020glu} to enhance feature learning and then adding the residual connection:
\begin{align}
\mathbf{x}^{l} &= \text{FFN}_l (\text{RMSN}(\mathbf{o}^{l})) + \mathbf{o}^{l} \\
\text{FFN}_l(\mathbf{x}) &= \text{SwiGLU}(\mathbf{x})\mathbf{W}_3^l = (\phi(\mathbf{x} \mathbf{W}_1^l) \otimes (\mathbf{x} \mathbf{W}_2^l)) \mathbf{W}_3^l
\end{align}
where $\phi$ represents SiLU, $\otimes$ denotes element-wise multiplication, and $\mathbf{W}_1^l \in \mathbb{R}^{d\times d_{FFN}}, \mathbf{W}_2^l \in \mathbb{R}^{d\times d_{FFN}}, \mathbf{W}_3^l \in \mathbb{R}^{d_{FFN}\times d}$ are learnable parameters. This configuration allows the network to effectively capture complex interactions within the data while maintaining efficient gradient flow through the residual connections.

\subsection{Prediction Layer \& Optimization objective}

After passing through $b$ layers of \textit{FuXi} blocks, each position has obtained sufficient information about the previously interacted items. We employ a multiplication with the transpose of the input embedding matrix, followed by a softmax function to obtain a probability distribution over predicted items. The transformation can be mathematically represented as follows:
\begin{shrinkeq}{-5px}
\begin{align}
    P\left(i_{t}^{(u)} = i \mid i_1^{(u)}, \dots, i_{t-1}^{(u)} \right) = softmax\left (\mathbf x^{b} \mathbf E^T   \right)_i
\end{align}
\end{shrinkeq}
In order to accelerate the training process, we adopt the sampled softmax loss with $N$ randomly sampled negative samples \cite{Klenitskiy_2023}.

\section{ANALYSIS}

\subsection{Space and Time Complexity}

\textbf{Space Complexity} 
Each \textit{FuXi} block comprises an AMS layer and an MFFN. The AMS layer features four projection matrices totaling $6d \times d_H$ parameters, alongside positional and temporal embeddings with $O(n + n_B)$ parameters, where $n_B$ is the number of buckets. The MFFN includes four projection matrices, amounting to $3d_h \times d + 3d_{FFN} \times d$ parameters. The item embeddings have $|\mathcal I| \times d$ parameters. Typically, $d_h$ and $d_{FFN}$ are proportional to $d$, and $n$ is comparable to $n_B$. Therefore, we assume $d_h = O(d)$, $d_{FFN} = O(d)$, and $n_B = O(n)$. \textit{FuXi}-$\alpha$ is formed by stacking $b$ \textit{FuXi} layers, leading to a total space complexity of $O(b(d^2 + n) + |\mathcal I|d)$.

\textbf{Time Complexity} 
The time complexity for computing attention weights in the semantics channel is $O(n^2d)$, compared to $O(n^2)$ in other channels. Calculating the QKV matrices and the MFFN both require $O(nd^2)$. The cost for generating predictions is $O(n|\mathcal I|d)$. Thus, the overall time complexity is $O(bn^2d+n(bd^2+|\mathcal I|d))$.

\subsection{Polynomial Approximation}

Next, we examine the properties of explicit inter-item interactions implemented by \textit{FuXi}-$\alpha$. To better analyze these interactions, we simplify the $l$-th layer of the \textit{FuXi} Block by treating attention weights as constants and omitting the second stage of the MFFN, activation functions, and most projection transformations. This simplification yields:
\begin{shrinkeq}{-5px}
\begin{align}
f_{block}^{(l)}(x_i; x_1, \cdots x_n) = x_i \circ \left( \sum_{j=1}^n a_{i,j}^{(l)}x_j \right) + x_i
\end{align}
\end{shrinkeq}
where the vectors $x_1, \ldots, x_n$ are the latent representations input to the $l$-th layer of the \textit{FuXi} block; $\circ$ denotes the interaction operator, such as element-wise multiplication; and $a_{i,j}^{(l)}$ are the attention weights in the $l$-th layer. In this section, let $x_{l, i}$ denote the output latent representation of the $i$-th item after the $l$-th layer. Let $F_n$ denote a polynomial of the form $\sum_{\boldsymbol{\alpha}} w_{\boldsymbol{\alpha}} \prod_{i} x_{0,i}^{\alpha_i}$, where the sum includes all terms satisfying $\sum \alpha_i \leq n$. We will use mathematical induction to show that $x_{b, i} = x_{i, 0} F_{2^b - 1}$.
\subsubsection{Base Case} Consider $b = 0$. Here, $x_{0,i} = x_{0,i} \cdot 1 = x_{0,i} \cdot F_0$, confirming the equation holds.
\subsubsection{Inductive Step} Assume the property holds for some integer $l \geq 0$. Now consider $b = l + 1$:
\begin{shrinkeq}{-3px}
\begin{align}
  x_{l+1, i} &= x_{l, i} \circ \sum_{j = 1}^n a_{i, j}^{(l+1)} x_{l, j} + x_{l, i} \\
  &= x_{0, i}F_{2^{l}-1} \circ \left (\sum_{j=1}^na_{i, j}^{(l+1)} x_{0, j} F_{2^{l}-1} + 1\right)
\end{align}
\end{shrinkeq}
For any term of the form $\prod_j x_{0,j}^{\alpha_i}$, where $1 \leq \sum \alpha_i \leq 2^{l+1}$, it appears in the expression $\sum_{j=1}^n a_{i,j}^{(l+1)} x_{0,j} F_{2^l-1}$. Thus, we have
\begin{shrinkeq}{-3px}
    \begin{align}
      \sum_{j=1}^n a_{i,j}^{(l+1)} x_{0,j} F_{2^l-1} + 1 = F_{2^l} 
    \end{align}
\end{shrinkeq}
Therefore, it follows that $x_{l+1,i} = x_{0,i} F_{2^{l+1} - 1}$.

Consequently, after progressing through $b$ layers of the \textit{FuXi} blocks, $x_{b, i}$ incorporates the outcome of feature interaction between $x_{0, i}$ and the result of interactions among all the items being of any degree up to $2^l - 1$.

\subsection{Analysis of AMS}

 The formulation of relative positional embeddings in the T5 architecture \cite{raffel2020t5} is delineated as follows.
The attention weights $\mathbf A = (a_{i, j})_{n\times n}$ can be computed by the process:
\begin{shrinkeq}{-5px}
    \begin{align}
        \mathbf A = \phi\left((\mathbf x \mathbf{W}_q)(\mathbf x \mathbf W_k)^T + \mathbf{B}\right)
    \end{align}
\end{shrinkeq}
where $\phi$ denotes a non-linear function, such as softmax or SiLU, and $\mathbf B = (b_{i, j})_{n\times n}$ denotes the matrix of the relative positional bias term. Let $q_i \in \mathbb R^{1 \times n}$ denotes the query vector of the $i$-th item, and $k_i, v_i, u_i$ denotes the key vector, the value vector, the vector used for Hadamard product respectively. 
The output of multi-head self-attention $o_i$ of the $i$-th item is then computed as:
\begin{shrinkeq}{-5px}
\begin{align}
    o_i &= W_o\left (\left (\sum a_{i,j} v_j \right)\otimes u_i \right) \\
    &\approx W_o \left (\left (\sum \phi_1(q_i k_j^T) V_j \right) \otimes u_i\right) +  W_o \left (\left (\sum \phi_2(b_{i,j}) v_j \right) \otimes u_i \right)
\end{align}
\end{shrinkeq}

On the other hand, in the AMS layer, the calculation process
is expressed as:
\begin{shrinkeq}{-5px}
\begin{align}
    o_i &= W_{o1}\left (\left (\sum \phi(q_i k_j^T) V_j \right) \otimes u_{i}^{(1)}\right) +  W_{o2} \left (\left(\sum b_{i,j} V_j \right) \otimes u_{i}^{(2)} \right)
\end{align}
\end{shrinkeq}
where $W_{o1}, W_{o2}$ denote the parameters in the first stage of the MFFN, and vectors $u_{i}^{(1)}$ and $u_{i}^{(2)}$ correspond to the $u_i$ vectors within the semantics and positional channels, respectively.
This demonstrates that the AMS layer facilitates a more expressive representation of positional and temporal information compared to the direct addition of attention weights, suggesting an enhancement in the model's capacity to leverage the temporal and positional information.

\subsection{Relationship with Existing Models}

Our work shares structural similarities with three models: SASRec \cite{kang2018self}, LLaMa \cite{dubey2024llama}, and HSTU \cite{zhai2024actions}. Here, we highlight the key differences between these models and our approach.

\subsubsection{SASRec and LLaMa} Unlike SASRec and LLaMa, which employ standard NLP architectures for recommendation systems, our model introduces two major innovations. First, instead of the traditional multi-head self-attention layer, we use the AMS layer to independently model temporal, positional, and semantic information, improving the model's feature utilization. Second, our model incorporates the MFFN, diverging from the FFN used in SASRec and LLaMa, by processing multi-channel information from the self-attention layer and enabling implicit feature interaction.

\subsubsection{HSTU} HSTU incorporates relative temporal and positional data by adding these features directly to attention weights, which can dilute their impact. Moreover, HSTU lacks an FFN layer, relying solely on self-attention and explicit feature interactions, limiting its ability to capture complex item relationships. Our model overcomes these limitations by decoupling temporal, positional, and semantic information within the self-attention layer and leveraging the MFFN to facilitate implicit interactions.

\section{EXPERIMENTS}\label{experiment}
\begin{table}[t]
\setlength{\abovecaptionskip}{-0cm}
\setlength{\belowcaptionskip}{-0.1cm}
 \caption{\small{Dataset statistics.}}
 \centering
 	\setlength{\tabcolsep}{1mm}
 \begin{tabular}{@{} c|c|c|c|c @{}}
 \hline
 \textbf{Dataset} 	  & \textbf{User}   & \textbf{Item} & \textbf{Interactions} & \textbf{Avg. Len.}   \\
 \hline
 MovieLens-1M  & 6,041 & 3,706 & 1,000,209 & 165.60   \\
 MovieLens-20M & 138,493 & 26,744 & 20,000,263 & 144.41   \\
 KuaiRand & 25,634 & 7,550 & 6,945,823 & 270.96 \\ 
 Industrial & 19,252,028 & 234,488 & 1,023,711,774 & 53.17 \\ 
 \hline
\end{tabular}
\label{tab:dataset_statistics}
\vspace{-5mm}
\end{table}

\begin{table*}[t]
\setlength{\abovecaptionskip}{0cm}
\setlength{\belowcaptionskip}{0cm}
\centering
\caption{The overall performance comparison. 
We use $\star$ to indicate a statistically significant result comparing FuXi-$\alpha$ with the best baseline which is indicated by underlined numbers.}
\setlength{\tabcolsep}{1mm}{
\small
\begin{tabular}{c|c|c|c|c|c|c|c|c|c|c|c|c|c|c|c}
\midrule[0.25ex]
Dataset &
\multicolumn{5}{c|}{MovieLens-1M} & 
\multicolumn{5}{c|}{MovieLens-20M} &
\multicolumn{5}{c}{KuaiRand} \\ \hline 
\textbf{Model} & NG@10 & NG@50 & HR@10 & HR@50 & MRR &  NG@10 & NG@50 & HR@10 & HR@50 & MRR & NG@10 & NG@50 & HR@10 & HR@50 & MRR \\\hline \hline
\textbf{BPRMF} & 0.0607 & 0.1027 & 0.1185 & 0.3127 &  0.0556 & 0.0629 & 0.1074 & 0.1241 & 0.3300 & 0.0572 & 0.0248 & 0.0468 & 0.0520 & 0.1560 & 0.0235 \\
\textbf{GRU4Rec} & 0.1015 & 0.1460 & 0.1816 & 0.3864 & 0.0895 & 0.0768 & 0.1155 & 0.1394 & 0.3177 & 0.0689 & 0.0289 & 0.0531 & 0.0597 & 0.1726 & 0.0275 \\
\textbf{NARM} & 0.1350 & 0.1894 & 0.2445 & 0.4915 & 0.1165 & 0.1037 & 0.1552 & 0.1926 & 0.4281 & 0.0910 & 0.0411 & 0.0747 & 0.0836 & 0.2399 & 0.0387\\

\hline
\textbf{SASRec} & 0.1594 & 0.2187 &0.2824 & 0.5500 & 0.1375 & 0.1553 & 0.2119 & 0.2781 & 0.5353 & 0.1330 & 0.0486 & 0.0877 &  0.0978 & 0.2801  & 0.0454\\
\textbf{LLaMa} & 0.1620 & 0.2207 & 0.2926 & 0.5591 & 0.1373 & 0.1640 & 0.2206& 0.2915&0.5476 & 0.1402 & 0.0495 & 0.0878 & 0.0973  & 0.2752 & 0.0466\\
\textbf{HSTU} & 0.1639 & 0.2238 & 0.2969 & 0.5672 & 0.1390 & 0.1642 & 0.2225 & 0.2909 & 0.5553 & 0.1410 & 0.0491 & 0.0861 & 0.0992 & 0.2718 & 0.0451 \\
\textbf{FuXi-$\alpha$} & 0.1835 & 0.2429 & 0.3254 & 0.5941 &0.1557 & 0.1954 & 0.2533 & 0.3353 & 0.5969 & 0.1677 & 0.0537 & 0.0942 & 0.1067  & 0.2951 & 0.0497 \\
\hline 
\textbf{SASRec-Large} & 0.1186 & 0.1733 & 0.2183 & 0.4671 & 0.0186 & 0.0206 & 0.0379 & 0.0412 & 0.1209 & 0.0207 & 0.0285 & 0.0428 & 0.0544 & 0.1227  & 0.0258 \\
\textbf{LLaMa-Large} & 0.1659 & 0.2257 & 0.2990 & 0.5692 & 0.1408 & 0.1842 & 0.2412 & 0.3202 & 0.5776 & 0.1576 & 0.0494 & 0.0878 & 0.0970 & 0.2754 & 0.0466\\
\textbf{HSTU-Large} & \underline{0.1844} & \underline{0.2437} & \underline{0.3255} & \underline{0.5929} & \underline{0.1568}  & \underline{0.1995} & \underline{0.2572} & \underline{0.3407} & \underline{0.6012} & \underline{0.1714}  & \underline{0.0494} & \underline{0.0883} & \underline{0.0990} & \underline{0.2799} & \underline{0.0460}  \\
\textbf{FuXi-$\alpha$-Large}  & \textbf{0.1934} &  \textbf{0.2518} & \textbf{0.3359} & \textbf{0.5983} & \textbf{0.1651} & \textbf{0.2086} & \textbf{0.2658} & \textbf{0.3530} & \textbf{0.6113} & \textbf{0.1792} & \textbf{0.0555} & \textbf{0.0963} & \textbf{0.1105} & \textbf{0.2995} & \textbf{0.0510}  \\ \hline
\end{tabular}
}
\label{tab:public_performance}
\end{table*}

\begin{table}[t]
    \setlength{\abovecaptionskip}{-0.0cm}
    \setlength{\belowcaptionskip}{-0.2cm}
    \centering
    \caption{Performance comparison on Industrial dataset.}
    \setlength{\tabcolsep}{1mm}{
    \small
    \begin{tabular}{c|c|c|c|c|c}
    \midrule[0.25ex]
    Dataset &
    \multicolumn{5}{c}{Industrial}  \\ \hline 
    \textbf{Model} & NG@10 & NG@50 & HR@10 & HR@50 & MRR \\\hline \hline
    
    \textbf{SASRec} & 0.1009 & 0.1580 & 0.1970 & 0.4581 & 0.0868 \\
    \textbf{LLaMa} & 0.1681 & 0.2238 & 0.2985 & 0.5498 & 0.1426 \\
    \textbf{HSTU} & \underline{0.1733} & \underline{0.2289} & \underline{0.3057} & \underline{0.5565} & \underline{0.1472}  \\
    \textbf{FuXi-$\alpha$}  & \textbf{0.1875} & \textbf{0.2424} & \textbf{0.3230} & \textbf{0.5702} & \textbf{0.1601}  \\ \hline
    \end{tabular}}
    \label{tab:industrial_performance}
\end{table}

\begin{table}[t]
    \setlength{\abovecaptionskip}{-0.0cm}
    \setlength{\belowcaptionskip}{-0.2cm}
    \centering
    \caption{Efficiency comparison on KuaiRand dataset with different sequence length.}
    \setlength{\tabcolsep}{1mm}{
    \small
    \begin{tabular}{c|c|c|c|c}
    \midrule[0.25ex]
     Dataset &
    \multicolumn{4}{c}{KuaiRand}  \\ \hline 
    \textbf{Model} & TPS@200 & TPS@400 & TPS@600 & TPS@800  \\\hline \hline
    
    \textbf{SASRec} & 2481 & 2024 & 1672 & 1398 \\
    \textbf{LLaMa} & 2330 & 1972 & 1602 & 1326 \\
    \textbf{HSTU} & 2078 & 1183  & 680 & 436  \\
    \textbf{FuXi-$\alpha$} & 1971 & 1053 & 615 & 394  \\ \hline
    \end{tabular}}
    \label{tab:efficiency}
\end{table}

\begin{table}[t]
    \setlength{\abovecaptionskip}{-0.0cm}
    \setlength{\belowcaptionskip}{-0.2cm}
    \caption{Performances of different FuXi-$\alpha$ variants.}
        \centering
    \setlength{\tabcolsep}{1mm}{
    \small
    \begin{tabular}{c|c|c|c|c|c|c}
    \midrule[0.25ex]
    Dataset &
    \multicolumn{2}{c|}{MovieLens-1M} & 
    \multicolumn{2}{c|}{MovieLens-20M} &
    \multicolumn{2}{c}{KuaiRand} \\ \hline 
    Model & NG@10 & HR@10 &  NG@10 & HR@10  & NG@10 & HR@10 \\\hline \hline
    \textbf{Base} & 0.1454 & 0.2676  & 0.1452 & 0.2647  & 0.0476 & 0.0928     \\
    \textbf{w/o AMS}  & 0.1563 & 0.2847 & 0.1612 & 0.2888 & 0.0470 & 0.0921      \\
    \textbf{w/o MFFN} & 0.1878 & 0.3304  & 0.2056 & 0.3488 & 0.0534 & 0.0947  \\
    \textbf{FuXi-$\alpha$}  & 0.1934  & 0.3359  & 0.2086 & 0.3530 & 0.0555 & 0.1105  \\
    \hline \hline
    \end{tabular}}
    \vspace{-3mm}
    \label{tab:Impact of main parts}
\end{table}


\subsection{Experiment Setup}\label{ExperimentSetup}

\subsubsection{Datasets}

To evaluate the performance of the proposed \textit{FuXi}-$\alpha$ architecture, we conduct extensive experiments on four real-world datasets, including three public datasets and one private large-scale dataset, which are described as follows:

\begin{itemize}[leftmargin=*,align=left]
  \item \textbf{MovieLens-1M} and \textbf{MovieLens-20M}\footnote{https://grouplens.org/datasets/movielens/}. 
    The MovieLens dataset is a widely used movie recommendation dataset, which contains users' rating and tagging activities.
    It has multiple subsets of different sizes.
    We select two subsets, MovieLens-1M and MovieLens-20M for our experiments.
  \item \textbf{KuaiRand}\footnote{https://kuairand.com/}. 
    This dataset is collected with the user logs of a video-sharing app from kuaishou.
    Users in this platform are usualy very active, with more than 200 interactions on average.
  \item \textbf{Industrial} 
    This dataset is constructed from user records of a mainstream music listening app, which has tens of millions active users every month.
    We construct users' behavior sequence with over a month of positive behaviors, including collect, like, play and so on.
\end{itemize}
For the first two datasets (\textbf{MovieLens-1M} and \textbf{MovieLens-20M}), we use the pre-processed train/validation/test set \footnote{https://github.com/facebookresearch/generative-recommenders} as in HSTU \cite{zhai2024actions} from Meta exactly.
For the latter two datasets (\textbf{KuaiRand} and \textbf{Industrial}), we process them using a similar manner to HSTU \cite{zhai2024actions} by ourself.
The statistics are shown in Table \ref{tab:dataset_statistics}.

\subsubsection{Compared Baseline}
For a comprehensive comparison, we compare \textit{FuXi}-$\alpha$ against two types of representative baselines: 
i) conventional models, including BPRMF \cite{rendle2012bpr}, GRU4Rec \cite{hidasi2015session}, and NARM \cite{li2017neural}; ii) autoregressive generative models, including SASRec \cite{kang2018self}, LLaMa \cite{dubey2024llama}, and HSTU \cite{zhai2024actions}.

\subsubsection{Evaluation Metrics}
We employ the widely used top-K Hit Ratio (HR@$K$), Normalized Discounted Cumulative
Gain (NDCG@$K$) and Mean Reciprocal Rank (MRR) to evaluate the recall performances.
For all metrics, higher value means better performance. 
We rank the ground-truth item from full set of items and report the performance of $K = 10, 50$ by default.

\subsubsection{Parameter Settings}
We implement our proposed \textit{FuXi}-$\alpha$ with Pytorch \cite{paszke2019pytorch}.
To enable large-scale model training, we apply the multi-machine and multi-card parallelism with the Accelerate library \cite{kotter2012accelerate}.
For a fair comparison, we maintain the same model parameters as HSTU \cite{zhai2024actions} in the first two datasets,
except for the number of layers.
For the KuaiRand dataset, we set the hidden dimension as 50, and the number of negative samples as 128 by default.
All other parameters like optimizer, learning rate and weight decay are consistent with HSTU \cite{zhai2024actions}.
For all the three datasets, the embedding dimensions and self-attention hidden vector dimensions are identical.
For the basic modeling capacity comparison, we set the number of layers as 2.
We also extend these generative models to deeper layers by stacking 4x number of layers (8 layers) and denoting it as "XX-Large" to analyze scaling effects.

\subsection{Performance Comparison  (RQ1)}\label{PerformanceComparison}
\subsubsection{Public Dataset Performance}
The overall performance comparison of the proposed \textit{FuXi}-$\alpha$ and baseline models are shown in Table \ref{tab:public_performance}.
Based on the results, we have the following observations:

\begin{itemize}[leftmargin=*,align=left]
    \item Firstly, the generative models (i.e., SASRec, LLaMa, HSTU, and \textit{FuXi}-$\alpha$) outperform conventional models (i.e., BPRMF, GRU4Rec and NARM), 
    even when equipped with just two layers of parameters.
    This demonstrates the generative models' superior ability in capturing complex item relationships and diverse user preferences by their autoregressive modeling paradigm.
    
    \item Secondly, as an early sequential model, SASRec fails to scale up to 8 layers across all three datasets, with a significant performance drop when the number of layers is increased to 8.
    In contrast, the two recently proposed transformer-based architectures, LLaMa and HSTU, show substantial improvements in the first two datasets.
  
    \item Finally, \textit{FuXi}-$\alpha$ consistently obtains the best results on all three datasets with all evaluation metrics, no matter it's a shallow network or a deep network.
    This demonstrates the outstanding ability of our proposed \textit{FuXi}-$\alpha$.
    Specifically, for shallow network, it outperforms the strongest baseline HSTU by 13.24\% in NDCG@10 (10.59\% in NDCG@50, 10.81\% in HR@10, 6.94\% in HR@50, 13.72\% in MRR) on average of the three datasets.
    For deep network, it outperforms the strongest baseline HSTU-Large by 7.26\% in NDCG@10 (5.24\% in NDCG@50, 6.14\% in HR@10, 3.19\% in HR@50, 6.90\% in MRR) on average of the three datasets.
    The excellent performance of \textit{FuXi}-$\alpha$ demonstrates the great utility of introducing explicit and implicit feature interaction for dedicated user behavior modeling.

\end{itemize}

\subsubsection{Industrial Dataset Performance}
Table \ref{tab:industrial_performance} presents the performance comparison of our proposed \textit{FuXi}-$\alpha$ against several baseline models on a private, large-scale industrial dataset.
The current online baseline in this scenario is a multi-channel recall system, with SASRec as one of the channels that recalls items based on embedding similarity. 
The music recalled from multiple channels is mixed together and then passed through a cascaded pre-ranking and ranking process to obtain the final recommended music list.
From Table \ref{tab:industrial_performance}, we have two key observations.
Firstly, the newly proposed LLaMa and HSTU significantly outperform SASRec in this music recommendation scenario, achieving gains of 64.82\% and 71.75\% in NDCG@10, respectively.
Secondly, our \textit{FuXi}-$\alpha$ outperforms both LLaMa and HSTU by 11.54\% and 8.19\%, respectively.
These substantial improvements highlight the potential of scaling laws, and the superiority of our proposed \textit{FuXi}-$\alpha$.

\begin{figure}
    \centering
    \setlength{\abovecaptionskip}{0pt}
    \setlength{\belowcaptionskip}{-15pt}
        \includegraphics[width=0.8\linewidth]{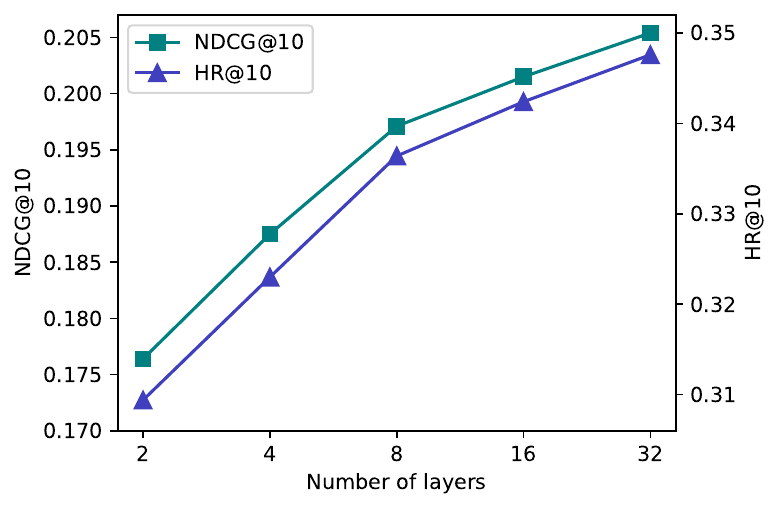}
    \caption{Scaling of \textit{FuXi}-$\alpha$ on Industrial Dataset.}
    \label{fig:industrial_scaling}
\end{figure}

\subsubsection{Scaling of \textit{FuXi}-$\alpha$ on Industrial Dataset}
Figure~\ref{fig:industrial_scaling} presents the performance of our proposed \textit{FuXi}-$\alpha$ on the industrial dataset when scaling up the number of layers while keeping all other hyper-parameters unchanged. 
Due to the memory limitation, we only scale up the layers to 32.
We observe that \textit{FuXi}-$\alpha$ adheres to the scaling law, as the results show a positive relationship between the model's performance and its size.
This is a highly attractive property as the performance can be further improved by scaling up of the model size, its training data, and the computational resources used.

\subsection{Efficiency Comparison (RQ2)}\label{EfficiencyComparison}

We assess the efficiency of the \textit{FuXi}-$\alpha$ architecture by comparing its Throughput Per Second (TPS) with generative baseline models. Experiments were conducted on the KuaiRand dataset with sequence lengths ranging from 200 to 800. Each experiment involved three complete forward and backward propagations across the dataset, calculating the average number of training samples processed per second. All hyperparameters, except sequence length, were consistent with previous experiments. Table \ref{tab:efficiency} shows the TPS results. As sequence length increases, TPS for all models decreases. Notably, SASRec and LLaMa outperform HSTU and \textit{FuXi}-$\alpha$ in TPS, likely due to their exclusion of temporal information encoding, which, while performance-enhancing, is time-intensive. Consequently, \textit{FuXi}-$\alpha$ achieves similar TPS to HSTU but significantly better overall performance, as seen in Tables \ref{tab:public_performance} and \ref{tab:industrial_performance}.

\subsection{Ablation Study (RQ3)}
 

To assess the effectiveness of sub-modules in our \textit{FuXi}-$\alpha$ architecture, we analyze three model variants: 
(1) \textbf{\textsl{Base Model}}: Replaces the AMS module with the vanilla self-attention layer from SASRec and substitutes the MFFN module with a single-stage MLP from HSTU. 
(2) \textbf{\textsl{w/o AMS}}: Replaces the AMS module with the vanilla self-attention layer. 
(3) \textbf{\textsl{w/o MFFN}}: Substitutes the MFFN module with a single-stage MLP.

Table \ref{tab:Impact of main parts} presents the ablation results, revealing the critical role of each component in model performance. Notably, removing the second stage of the MFFN results in a significant performance drop, emphasizing the importance of thorough implicit feature interactions. Despite this, the model still outperforms HSTU, demonstrating the effectiveness of our approach in capturing temporal and positional information. Additionally, replacing the AMS with the vanilla self-attention layer leads to a marked performance decline, highlighting the necessity of explicit feature interactions and effective use of temporal and positional data in recommendation tasks. These results confirm the essential contributions of each module to the model's predictive capability.

\subsection{Hyperparameter Study (RQ4)}

\begin{figure}[t]
    \centering    
    \setlength{\abovecaptionskip}{0pt}
    \setlength{\belowcaptionskip}{-5pt}
    \subfigure[MovieLens-1M] {
     \label{fig:a}     
    \includegraphics[width=0.47\columnwidth]{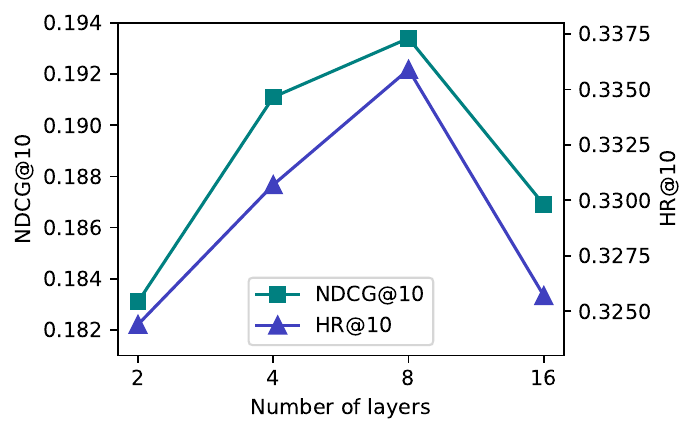}  
    }     
    \subfigure[KuaiRand] { 
    \label{fig:b}     
    \includegraphics[width=0.47\columnwidth]{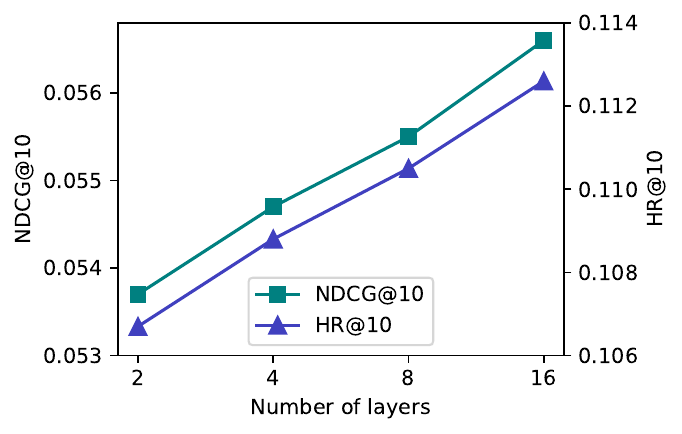}     
    }    
    \caption{Performances with different number of layers.}     
    \label{fig:layer}   
\end{figure}

\begin{figure}[t]
    \centering    
    \setlength{\abovecaptionskip}{0mm}
    \setlength{\belowcaptionskip}{-10pt}
    \subfigure[MovieLens-1M] {
     \label{fig:a}     
    \includegraphics[width=0.47\columnwidth]{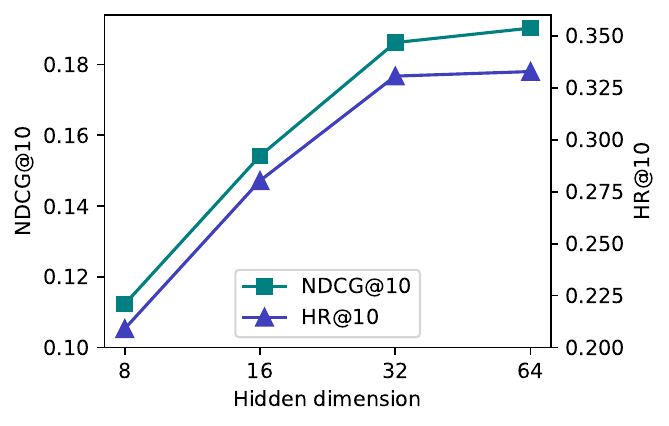}  
    }     
    \subfigure[KuaiRand] { 
    \label{fig:b}     
    \includegraphics[width=0.47\columnwidth]{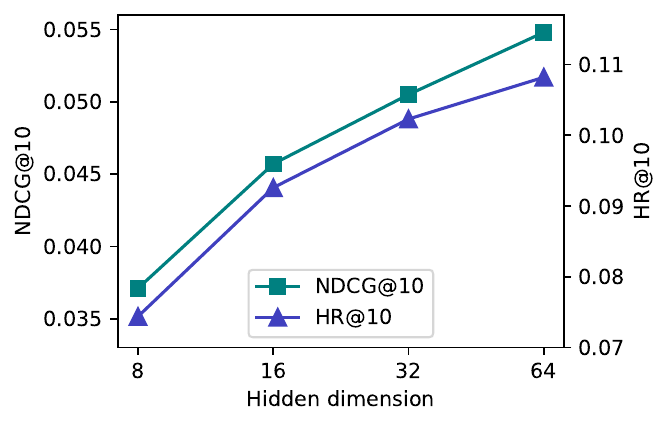}     
    }    
    \caption{Performances with different hidden dimension.}     
    \label{fig:dimension}     
\end{figure}

\begin{figure}[ht]
\centering
\setlength{\abovecaptionskip}{0pt}
\setlength{\belowcaptionskip}{0pt}
\subfigure[MovieLens-1M] {
 \label{fig:a}     
\includegraphics[width=0.47\columnwidth]{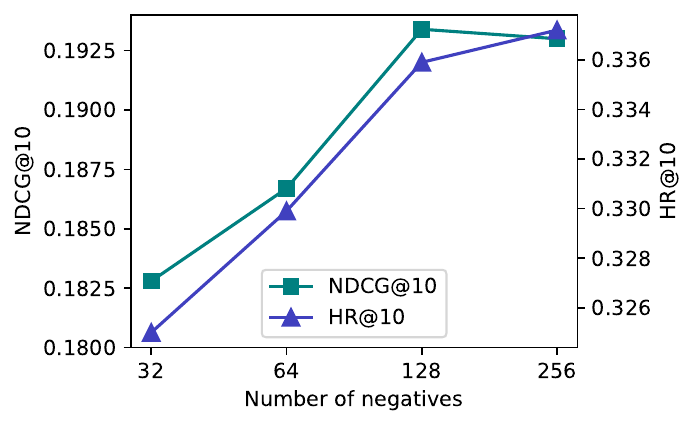}  
}     
\subfigure[KuaiRand] { 
\label{fig:b}     
\includegraphics[width=0.47\columnwidth]{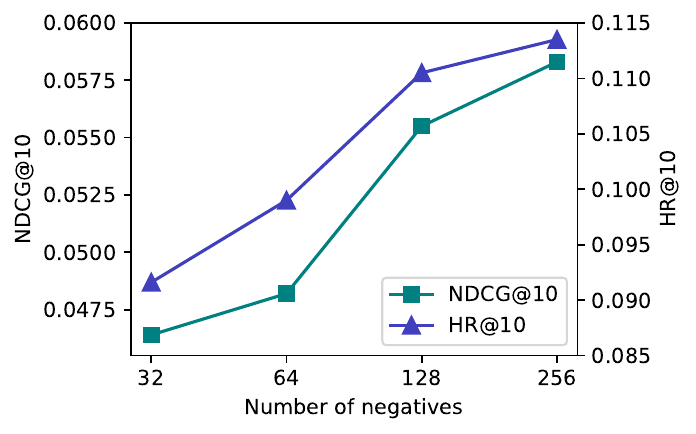}     
}    
\caption{Diverse negative sample counts in performances.
}
\label{fig:negative}
\end{figure}

We examine the effects of various hyper-parameters on \textit{FuXi}-$\alpha$, focusing on (1) the number of layers, (2) the hidden dimension, and (3) the number of negative samples for training. Due to space constraints, we present only NDCG@10 and HR@10 results for the MovieLens-1M and KuaiRand datasets. Results for other metrics (NDCG@50, HR@50, MRR) and datasets (MovieLens-20M) are similar but omitted. We alter one hyper-parameter at a time while keeping others constant to ensure fair comparisons.

\subsubsection{The number of layers}
Increasing layers is a rapid method to scale model parameters and enhance \textit{FuXi}-$\alpha$'s representational capacity. We vary layers from 2 to 16, as shown in Figure \ref{fig:layer}. On MovieLens-1M, performance improves from 2 to 8 layers, but declines at 16 layers. Conversely, on KuaiRand, performance consistently increases from 2 to 16 layers. This may be due to MovieLens-1M's smaller size limiting parameter scaling.

\subsubsection{The hidden dimension}
Uniform embedding and self-attention hidden dimensions are used across datasets. Increasing hidden dimensions enhances item representation and self-attention similarity accuracy. Adjusting dimensions from 8 to 64, Figure \ref{fig:dimension} shows performance on MovieLens-1M saturates at 32 dimensions, with minimal gains beyond. In contrast, KuaiRand performance steadily improves across all dimensions.

\subsubsection{Negative Samples}
The influence of negative sampling on large recommendation models has been overlooked in studies on LLM scaling laws \cite{kaplan2020scaling,hoffmann2022training}. We vary negative samples from 32 to 256, with results in Figure~\ref{fig:negative}. Performance improves on both datasets even beyond 64 negative samples, with gains from negative sampling surpassing those from layer increases. This underscores the critical role of negative sampling in enhancing 
models' performance.

\subsection{Online A/B Test}

In a main scenario of Huawei Music, we conducted a 7-day online A/B test to evaluate the performance of our new model, \textit{FuXi}-$\alpha$, utilizing 30\% of the user traffic. The results demonstrated that \textit{FuXi}-$\alpha$ achieved significant improvements compared to a well-optimized multi-channel retrieval baseline that has been refined over several years. Specifically, the average number of songs played per user increased by 4.67\%, while the average listening duration per user rose by 5.10\%. These findings indicate that \textit{FuXi}-$\alpha$ excels in enhancing user interaction and engagement, particularly by improving user experience and increasing platform usage time.
After evaluation of several weeks, the \textit{FuXi}-$\alpha$ had become an inherent channel in this scenario to serve most of the online traffic.


\section{CONCLUSION}

In our paper, we proposed a novel model called \textit{FuXi}-$\alpha$, which leverages Adaptive Multi-channel Self-attention to enhance the interactions with temporal and positional features, and Multi-stage Feed-Forward Networks (MFFNs) to facilitate implicit interactions. Our offline and online A/B experiments demonstrate that \textit{FuXi}-$\alpha$ consistently outperforms prior models, and reveal the effectiveness of each component. Additionally, the performance continually improves while scaling up our model, highlighting its potential for large-scale recommendation systems. In future work, we plan to extend our model to tackle more complex recommendation problems, such as multi-behavior and multi-modal recommendations, and to apply our model to scenarios involving long sequences.

\bibliographystyle{ACM-Reference-Format}
\bibliography{main}

\end{document}